\documentclass[twocolumn,prx,aps,floatfix,superscriptaddress,longbibliography]{revtex4-1}
\usepackage{lipsum} 
\usepackage[latin9]{inputenc}
\usepackage{amsmath}
\usepackage{amssymb}
\usepackage{graphicx}
\usepackage{esint}
\usepackage{times}
\usepackage{siunitx}
\usepackage{epsfig}
\usepackage{epstopdf}
\usepackage[table]{xcolor}
\usepackage{booktabs}
\usepackage{multirow}
\usepackage{float}
\usepackage{appendix}
\usepackage[colorlinks=true,
linkcolor=blue,
citecolor=blue,
urlcolor=blue]{hyperref}

\newcommand{\ZZU}{Quantum Information Institute, School of Physics and Laboratory of Zhongyuan Light, Zhengzhou University, Zhengzhou 450001, China}
\newcommand{\SXU}{State Key Laboratory of Quantum Optics and Quantum Optics Devices, Institute of Laser Spectroscopy, Shanxi University,Taiyuan 030006,China}
\newcommand{\HAS}{Institute of Quantum Materials and Physics, Henan Academy of Sciences, Zhengzhou 450046, China}
\newcommand{\NCU}{Department of Physics, Nanchang University, Nanchang 330031, China}

\begin{document}
	\title{Experimental Demonstration of a Measurement-Feedback Quantum Information Engine}
	
	\author{Jinfeng Wei}
	\altaffiliation{Co-first authors have the same contributions}	
	\affiliation{\ZZU} 
	
	\author{Yingying Hong}
	\altaffiliation{Co-first authors have the same contributions}
	\affiliation{\NCU}
	
	\author{Dehua Liu}
	\altaffiliation{Co-first authors have the same contributions}
	\affiliation{\NCU} 
	
	\author{Xi Wang}
	\affiliation{\ZZU} 
	
	\author{Zhe Wang}
	\affiliation{\ZZU} 
	\author{Yiling Zhan}
	\affiliation{\ZZU} 
	
	\author{Kaifeng Cui}
	\email{cuikaifeng@zzu.edu.cn}
	\affiliation{\ZZU} 
	
	\author{Jintao Bu}
	\affiliation{\ZZU} 
	
	\author{Jianhui Wang}
	\email{wangjianhui@ncu.edu.cn}
	\affiliation{\NCU}
	
	\author{Leilei Yan}
	\email{llyan@zzu.edu.cn}
	\affiliation{\ZZU}
	\affiliation{\HAS}
	
	\author{Gang Chen}
	\email{chengang971@163.com}
	\affiliation{\ZZU} 
	\affiliation{\SXU} 
	
	\begin{abstract}
		Harnessing finite-time nonadiabatic transitions that are conventionally associated with quantum inner friction for useful work extraction remains an open experimental challenge. Here we address this issue by introducing and experimentally realizing an innovative measurement-feedback quantum information engine model in the trapped $^{40}\mathrm{Ca}^{+}$ ion system, in which the projective measurement replaces the hot reservoir as a nonthermal energy source and the feedback control conditionally steers the system through either unitary compression-expansion strokes or thermalization. We experimentally show that the repeated feedback cycle converges to a stable operating regime with a resolved energetic balance, and, by controlling the measurement angle and stroke duration, measurement-induced coherence and the finite-time nonadiabatic contribution can enhance work extraction and raise the efficiency above the corresponding Otto benchmark. The experimental results further show that, over finite ranges of stroke durations, the efficiency and intrinsic cycle power can increase simultaneously. Our experiment establishes a route toward information-to-work quantum engines that convert finite-time irreversibility into performance-enhancing resources.  
	\end{abstract}
	\maketitle
	
	\textit{Introduction\text{---}}Maxwell's demon bridges information and thermodynamics by utilizing measurement information to extract work from a single heat bath~\cite{maxwell1871theory, szilard1929uber, leff1990maxwell}. This paradox is resolved by Landauer's principle and modern information thermodynamics, where feedback and mutual information modify the thermodynamic bounds~\cite{landauer1961irreversibility, Bennett1982, maruyama2009colloquium, parrondo2015thermodynamics, PhysRevLett.104.090602, Toyabe2010, PhysRevLett.113.030601}. As an archetype, quantum information engines exploit measurement records and backaction, together with feedback or adaptive control, to convert information and nonequilibrium resources into useful work~\cite{koski2015onchip, cottet2017observing, elouard2017quantum, PhysRevLett.120.260601, PhysRevE.105.044137, PhysRevLett.129.050401, PhysRevLett.130.240401, camati2016maxwell, Masuyama2018, PhysRevLett.121.030604}, including measurement-fueled implementations in which measurements that do not commute with the system Hamiltonian can inject energy into the working medium and replace a conventional thermal source~\cite{Elouard2017, PhysRevE.86.040106, PhysRevE.96.022108, PhysRevE.104.054128, PhysRevA.104.062210, PhysRevA.106.022436, rygc-bc3c,PhysRevE.110.L052102}. The development of measurement-based information engines has broadened the framework of quantum thermodynamics and deepened the understanding of information-to-work conversion, entropy production, and irreversibility~\cite{Goold_2016, PhysRevLett.127.040602, PhysRevE.106.054131}.
	
	To obtain a finite power output, information engines must operate within a finite duration, during which quantum inner friction arises from nonadiabatic transitions induced by the noncommutativity of the Hamiltonian at different times and is conventionally associated with irreversible work and degraded engine performance~\cite{PhysRevE.65.055102, PhysRevLett.113.260601, PhysRevE.68.016101, Singh2024, e22091060,PhysRevE.103.032144}. Thus, a common strategy for improving engine performance is to suppress quantum inner friction using control techniques~\cite{Science_sta, Berry_2009, Hou2025}, while quantum coherence has also been explored as a resource for enhancing thermodynamic performance~\cite{RevModPhys.89.041003, PhysRevLett.113.140401, PhysRevB.109.125407, Korzekwa_2016, PhysRevE.99.042105, PhysRevLett.122.110601, Kim2022, Lostaglio2015, Scully2011, PhysRevX.5.031044}. Although microscopic quantum heat engines have been experimentally demonstrated on several platforms~\cite{PhysRevLett.122.110601, Kim2022, single_atom_heat_engine, PhysRevLett.123.080602, PhysRevLett.123.240601, VanHorne2020}, the positive energetic role of quantum inner friction remains experimentally unresolved. Recently, some theoretical studies~\cite{PhysRevE.101.022127, Liu_2025} have revealed a different possibility: for suitable engine architectures and active input states, finite-time nonadiabatic transitions conventionally associated with inner friction can contribute positively to work extraction. However, such theoretical works remain limited, and the role of quantum inner friction has remained largely unexplored in quantum information engines; in particular, its deliberate use as a positive contribution to work has not been experimentally demonstrated. 
	
	\begin{figure*}[htbp]
		\centering
		\includegraphics[width=0.95\linewidth]{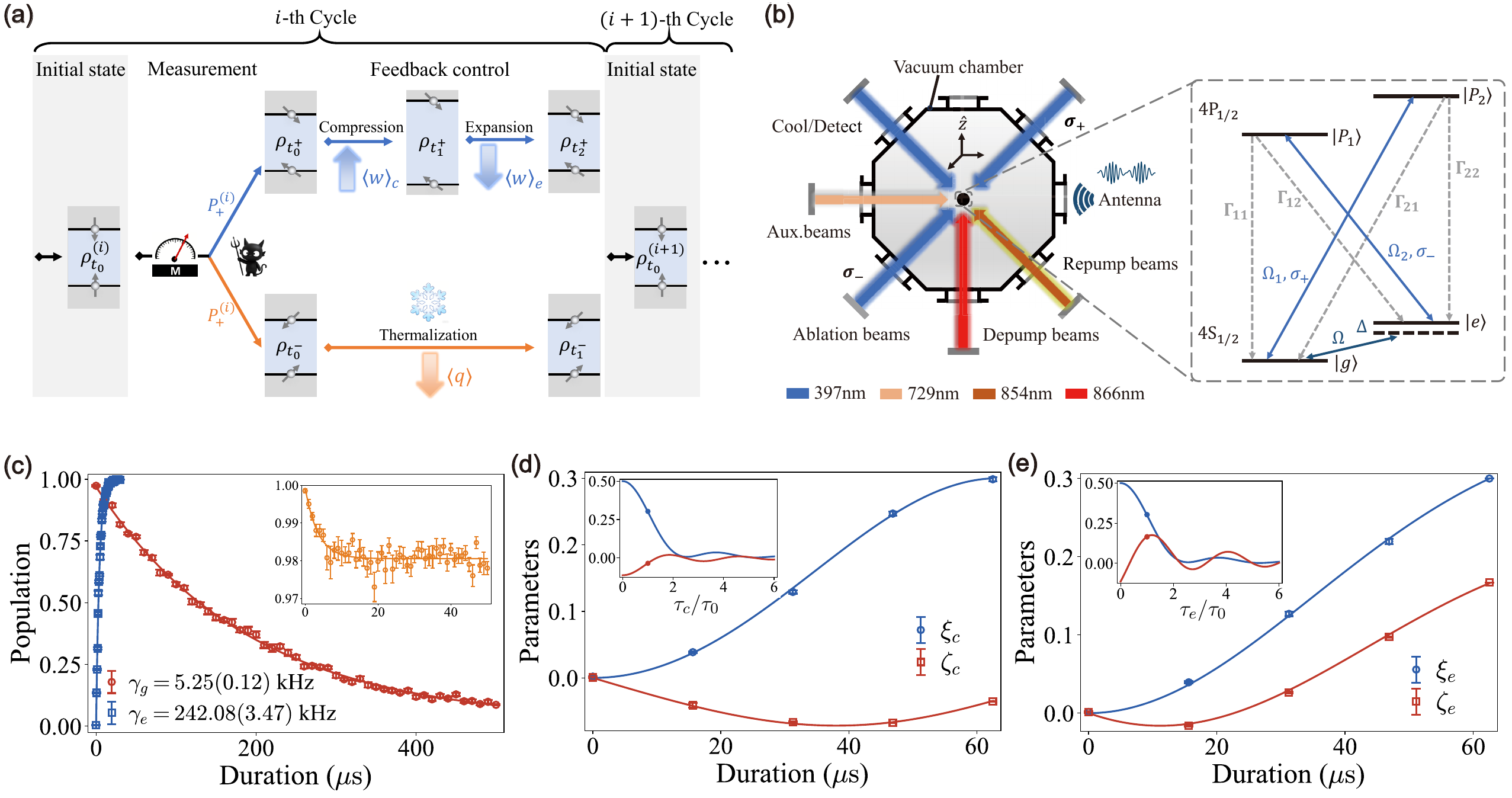}
		\caption{\textbf{Scheme and experimental set-up of MFQIE.} \textbf{(a)} Schematic of MFQIE in the two-level system, where the system starts from $\rho_{t_0}^{(i)}$ in the $i$th cycle, subsequently undergoes a measurement $M$ with the eigenvalues $a_{\pm}$ and corresponding eigenstates $|\psi_{\pm}\rangle$. If the outcome is $a_+$, the system collapses to the eigenstates $|\psi_+\rangle$ and is then controlled by a unitary compression-expansion stroke, while if the outcome is $a_-$, the system collapses to $|\psi_-\rangle$, and then undergoes a thermalization process. After the $i$th cycle, the state of the system evolves into $\rho_{t_0}^{(i+1)}$ and the next cycle starts. \textbf{(b)} The set-up of our experiment to realize the MFQIE, where an ion is trapped at the center of five-segment linear ion trap in a vacuum chamber~\cite{SM}, and illuminated by multiple different lasers or microwave beams to couple the different energy levels of the ion. \textbf{(c)} Effective dissipative process under different dissipation rates, where the inset shows the evolution of $|g\rangle$ under a dual-dissipation process, where the population of steady state is $P_g=\gamma_e/(\gamma_g+\gamma_e)=0.979(1)$, resulting in an effective temperature $T_c=0.26T_0$ with the temperature unit $T_0=\hbar\omega_c/k_B$. \textbf{(d)} and \textbf{(e)} The measured evolution of nonadiabatic-transition parameter $\xi$ and coherence-dependent dynamical parameter $\zeta$ for the compression-expansion process with durations $\tau_{c,e}=\tau_0$, respectively, where the insets show the final values of $\xi$ and $\zeta$ under different durations. Here, the data points and curves, also in the following figures, correspond to the experimental results and numerical simulations, respectively, and the parameters are selected as $\omega_c/2\pi = 4$~kHz, $\omega_h/2\pi = 8$~kHz and $\tau_{0}=\pi/2\omega_c$.}
		\label{fig1}
	\end{figure*} 
	
	Here, we touch this gap by experimentally realizing a measurement-feedback quantum information engine (MFQIE) in the trapped-ion system, where a projective measurement replaces the conventional hot reservoir of the Otto cycle, supplies nonthermal energy through measurement backaction, and simultaneously generates coherence in the energy eigenbasis, while feedback control based on the measurement outcome steers the system along either a finite-time unitary compression-expansion stroke or a thermalization stroke. By separately tuning the measurement basis and the unitary-stroke duration, we experimentally demonstrate that both the measurement-induced coherence and quantum inner friction in the finite-time unitary-stroke can increase work output and enhance efficiency beyond the corresponding Otto benchmark. In particular, over the experimentally explored ranges of stroke durations, the stroke-resolved quantum-inner-friction contribution remains nonnegative and enhances work extraction, while the efficiency and intrinsic cycle power can increase simultaneously, demonstrating that finite-time nonadiabatic dynamics can be harnessed to improve engine performance.
	
	\textit{Model\text{---}}Before presenting the experimental results, we briefly introduce the model of the MFQIE. As shown in \autoref{fig1}(a), the engine starts with an iterative initial state $\rho_{t_0}^{(i)}$ and Hamiltonian $H_0=\hbar\omega_c\sigma_z/2$ with fixed energy splitting $\omega_c$. Generally, the initial state can be an arbitrary state in the first cycle  and reaches the steady state $\rho_s$ after a finite number of cycles. Without loss of generality, we set $\rho_{t_0}^{(1)}=|g\rangle\langle g|$ with $|g\rangle$ and $|e\rangle$ denoting the lower and upper states of the system spin, respectively. Then, a projective measurement of observable $M(\theta)=\cos\theta\sigma_z+\sin\theta\sigma_x$ with $\theta$ denoting the measurement angle is applied and yields the outcomes $a_{\pm}=\pm 1$, thereby collapsing system to the corresponding eigenstates $|\psi_{\pm}\rangle$ with probabilities $p_\pm^{(1)}=\langle\psi_{\pm}|\rho_{t_0}^{(1)}|\psi_{\pm}\rangle$~\cite{SM}. Based on the outcomes, feedback control is then implemented to conditionally drive the evolution of the system. In the present experiment, the measurement-conditioned feedback map is implemented at the ensemble level. Because direct fluorescence readout heats the ion, the measurement statistics are first acquired, after which the ion is recooled and reprepared in the corresponding postmeasurement state before the selected feedback branch is applied; see the Supplemental Material~\cite{SM}.
	
	\begin{figure*}[htbp]
		\centering
		\includegraphics[width=0.94\linewidth]{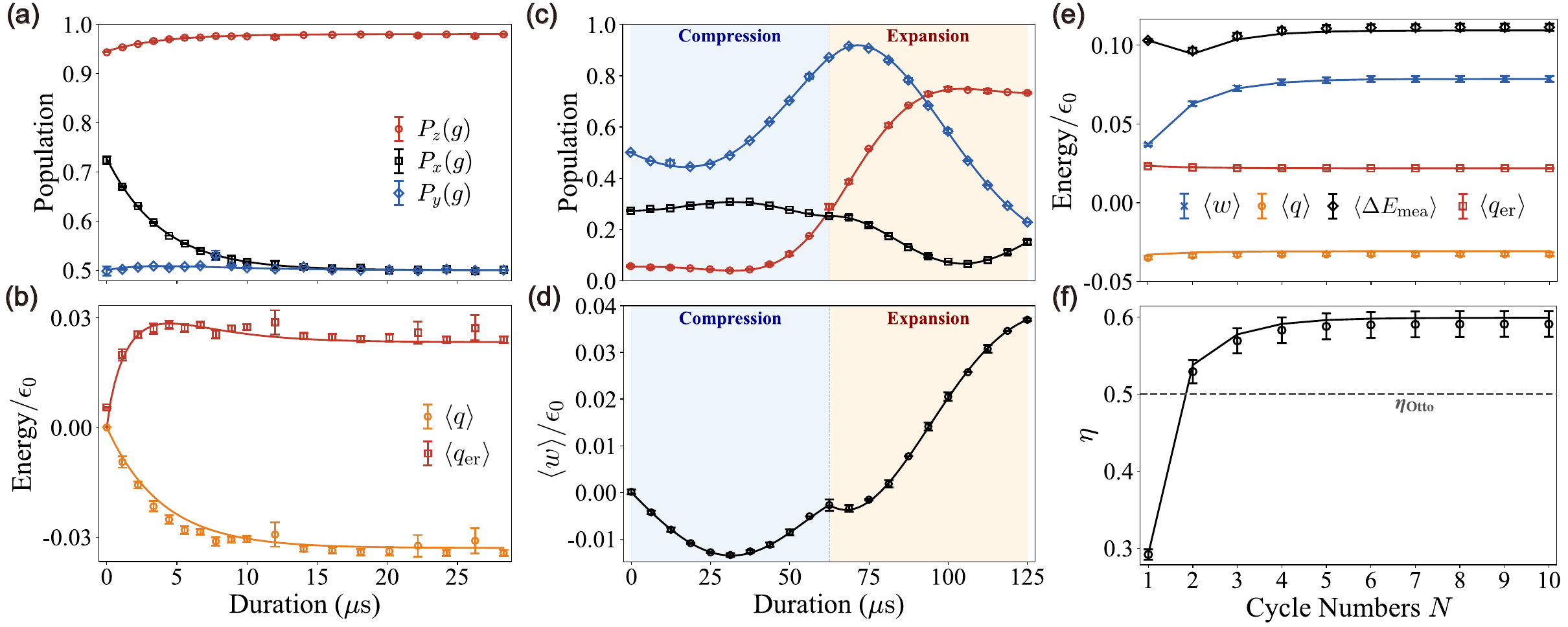}
		\caption{\textbf{Experimental demonstration of the MFQIE.} \textbf{(a)} and \textbf{(c)} The evolution of tomography measurement population of density matrix under the thermalization and unitary compression-expansion strokes, respectively. \textbf{(b)} and \textbf{(d)} The evolutions of heat exchange $q$ and erasure cost $\langle q_{\mathrm{er}}\rangle$ in the thermalization stroke, and extracted work $\langle w\rangle$ in the compression-expansion stroke, respectively. Panels (a-d) are obtained in the first cycle of engines. \textbf{(e)} Thermodynamic quantities and \textbf{(f)} engine efficiency for different cycles under a full thermalization duration, i.e., $\tau_{d} = 1/\sqrt{\gamma_g\gamma_e}$. Here the initial state of first cycle is $|g\rangle$, the dissipative rates are the same as those in \autoref{fig1}(c), the measurement angle $\theta=0.15\pi$, energy unit $\epsilon_0=\hbar\omega_c$, and the other parameters are the same as \autoref{fig1}.} 
		\label{fig2}
	\end{figure*}
	
	If the outcome is $a_-$, the system collapses to $|\psi_-\rangle$ at $t = t_0^-$ and is then coupled to a cold reservoir with the inverse temperature $\beta_c$  of the reservoir for a thermalization duration $\tau_{d}=t_1^--t_0^-$, during which the coherence is destroyed and the system evolves towards the thermal equilibrium state $\rho_{\rm th}=\exp(-\beta_c H_0)/Tr[\exp(-\beta_cH_0)]$. If the thermalization process is governed by the Lindblad master equation with the decay operators of $|g\rangle$ and $|e\rangle$ states corresponding to $L_g = \sqrt{\gamma_g}|e\rangle \langle g|$ and $L_e = \sqrt{\gamma_e}|g\rangle \langle e|$~\citep{SM}, the system gradually relaxes toward the steady state (i.e., thermal equilibrium state). The steady state and effective inverse temperature are given by $\rho_{\rm th}=(\gamma_g|e\rangle\langle e|+\gamma_e|g\rangle\langle g|)/(\gamma_g+\gamma_e)$ and  $\beta_c=(\hbar\omega_c)^{-1}\ln(\gamma_e/\gamma_g)$, respectively. 
	
	If the outcome is $a_+$, the system state undergoes the unitary compression-expansion control stroke. In the compression stage, the system is driven by a time-dependent Hamiltonian $H_c(t)=\frac{\hbar\omega(t)}{2}[\cos(\frac{\pi t}{2\tau_{c}})\sigma_z+\sin(\frac{\pi t}{2\tau_{c}})\sigma_x]$ for the duration $\tau_{c}=t_1^+-t_0^+$, where the energy splitting linearly increases as $\omega(t)=[\omega_h t-\omega_c(t-\tau_{c})]/\tau_{c}$~\cite{SM}. Whereafter, it undergoes a time-reversed process of the compression stage, i.e., the expansion stage, but with a different duration $\tau_{e}=t_2^+-t_1^+$. In this unitary process, the state successively evolves into $\rho_{t_1^+}=U_c\rho_{t_0^+}U_c^{\dagger}$ and $\rho_{t_2^+}=U_e\rho_{t_1^+}U_e^{\dagger}$ with $U_c$ ($U_e$) denoting the corresponding evolution operators of compression (expansion) stage~\citep{SM}. In this stroke, the quantum inner friction arises due to the non-commutativity of Hamiltonian at different times, and typically leads to irreversible nonadiabatic work, degrading the performance of the heat engine. However, we experimentally demonstrate that the nonadiabatic transitions generated during the finite-time unitary strokes can be controlled by adjusting the stroke durations to contribute positively to work extraction and efficiency. 
	
	To complete the thermodynamic accounting, we include a Landauer reset term associated with a state-bearing controller register, conditioned on the measurement outcome. This reset is not implemented as an additional physical stroke in the present experiment. The reset of the binary outcome record constitutes a distinct controller ledger, whose entropy-rate formulation is discussed in the companion theoretical work~\cite{Hong2026FeedbackEngine}. The cycle continues by repeating the above control protocol, and after a few cycles it reaches a stable cycle with fixed work, heat, and efficiency.
	
	\textit{System\text{---}}As shown in \autoref{fig1}(b), our experiment is implemented in a linear five-segment Paul trap, where the $^{40}\mathrm{Ca}^{+}$ ion is trapped at the trap center with the secular frequencies $(\omega_x, \omega_y, \omega_z)/2\pi=(1.89, 1.85, 0.36)$~MHz. The two-level system of the quantum engine is encoded in two Zeeman sub-levels of the ground state, defined as $|g\rangle=|4\mathrm{S}_{1/2}, m_s=-1/2\rangle$ and $|e\rangle=|4\mathrm{S}_{1/2}, m_s=1/2\rangle$ with an energy splitting of $\omega_s/2\pi=26.55$~MHz induced by a magnetic field of 9.485 G in the horizontal plane at an angle of $45^\circ$ relative to the $\hat{z}$ axis. A microwave with the controllable Rabi frequency and detuning is applied to couple the two-level system to execute the unitary compression-expansion stroke. The thermalization stroke is produced via two counter-propagating circularly polarized 397~nm laser beams, where the $\sigma_+$ ($\sigma_-$) laser couples $|g\rangle$ to $|P_2\rangle=|4\mathrm{P}_{1/2},m_s=1/2\rangle$ ($|e\rangle$ to $|P_1\rangle=|4\mathrm{P}_{1/2},m_s=-1/2\rangle$) with the Rabi frequency $\Omega_1$ ($\Omega_2$), yielding an effective dissipation rate $\gamma_g=\Omega_1^2/3\Gamma$ ($\gamma_e=\Omega_2^2/3\Gamma$) with $\Gamma$ denoting the natural linewidth of $P_{1/2}$. Due to the decay branches $\Gamma_{11}$ and $\Gamma_{22}$, two additional dephasing channels with strengths $2\gamma_g$ and $2\gamma_e$ are induced, which slightly speeds up thermalization of the system but have no significant effect on the results~\cite{SM}. As shown in \autoref{fig1}(c), the effective dissipation rates can be measured by fitting the population dynamics of initial states $|g\rangle$ and $|e\rangle$, respectively. 
	
	\begin{figure*}[htbp]
		\centering
		\includegraphics[width=0.96\linewidth]{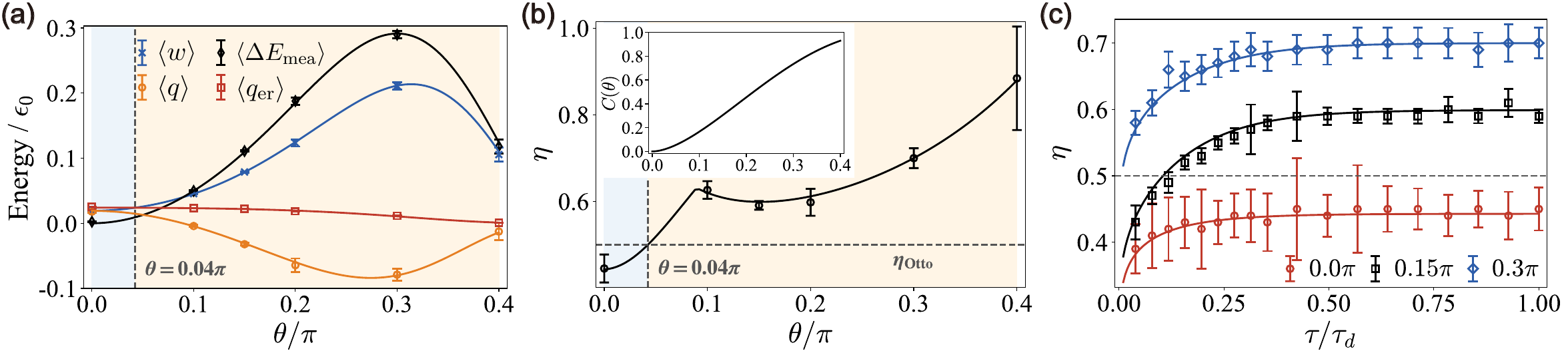}
		\caption{\textbf{Thermodynamic advantage of MFQIE in the stable cycle.} \textbf{(a)} The thermodynamic quantities and \textbf{(b)} engine efficiency advantage as a function of the measurement angle $\theta$ under full thermalization, where the inset in (b) shows the measurement-induced coherence. \textbf{(c)} The engine efficiency for different thermalization durations ($\tau$). The parameters are the same as \autoref{fig2}.}
		\label{fig3}
	\end{figure*}
	
	In the unitary compression-expansion stroke of MFQIE, the extracted work can be obtained as
	\begin{equation}
		\label{work}
		\langle w\rangle = p_+\epsilon_0[2(\zeta_{c}+\zeta_{e}-2\xi_{e}\zeta_{c})+(\xi_{c}+\xi_{e}-2\xi_{e}\xi_{c})\cos\theta],  
	\end{equation}
	where the adiabatic and coherence parameters are accordingly given as $\xi_{c}=|\langle e(t_1^+)|U_{c}|g(t_0^+)\rangle|^2$, $\xi_{e}=|\langle e(t_2^+)|U_{e}|g(t_1^+)\rangle|^2$, $\zeta_{c} = \text{Re}[U_{c}^{\text{gg}}\langle g(t_0^+)|\rho_{t_0^+}|e(t_0^+)\rangle {U_{c}^{\text{eg}}}^\dagger]$ and $\zeta_{e} = \text{Re}[U_{e}^{\text{gg}}\langle g(t_1^+)|\rho_{t_1^+}|e(t_1^+)\rangle {U_{e}^{\text{eg}}}^\dagger]$ with $U_{c,e}^{\text{gg,eg}}$ denoting the corresponding matrix elements of the evolution operator $U_{c,e}$~\cite{SM}. In \autoref{fig1} (d,e), we experimentally show the evolution of parameters $\xi$ and $\zeta$, where the adiabatic parameters are nonnegative while coherence parameters can be positive or negative depending on the durations. The quantum inner-friction contribution to the extracted work is then obtained as $\langle w\rangle_{\mathrm{fri}}=p_+\epsilon_0\cos\theta(\xi_c+\xi_e-2\xi_c\xi_e)$. Notably, for $|\theta|<\pi/2$, we will obtain an interesting result that the contribution of friction always remains nonnegative and enhances the performance of the engine whenever it is nonzero. In the adiabatic limit, the compression-expansion stroke quasistatically runs and results into $(\xi_{c,e}, \zeta_{c,e}\rightarrow 0)$ and zero average work output. 
	
	In the thermalization stroke, the average heat exchange and the working-medium entropy change are accordingly given by
	\begin{equation}
		\langle q\rangle=2^{-1} p_-\epsilon_0[\langle \sigma_z(t_1^-)\rangle+\cos(\theta)],
	\end{equation}
	and $\langle \Delta S\rangle=-p_-k_B\text{Tr}[\rho_{t_1^-}\ln\rho_{t_1^-}]$~\cite{SM}. By using the energy conservation, we can obtain the energy cost of measurement as $\langle \Delta E_{\text{mea}}\rangle=\langle w\rangle -\langle q\rangle$. Under this controller ledger, the minimum conditional reset charge is
	\begin{equation}
		\langle q_{\mathrm{er}}\rangle=-p_-\beta_c^{-1}\text{Tr}[\rho_{t_1^-}\ln\rho_{t_1^-}].
	\end{equation}
	The + branch gives no contribution to this term because its output is a unitary image of a pure postmeasurement state. Accordingly, within the thermodynamic ledger adopted here, the assigned input consists of the measurement-induced energy input, heat absorbed from the cold reservoir when present, and the conditional state-register reset charge, giving the operational efficiency as $\eta = \langle w\rangle/[\langle \Delta E_{\mathrm{mea}}\rangle+\mathcal{H}(\langle q\rangle)\langle q\rangle+\langle q_{\mathrm{er}}\rangle]$ with $\mathcal{H}(\cdot)$ denoting the Heaviside function. 
	
	\textit{Result\text{---}}In \autoref{fig2}, we experimentally demonstrate the cycle of MFQIE starting from the initial state $|g\rangle$, where the state evolves along two conditional branches, and the experimental results agree well with the numerical simulations. Fig.~\ref{fig2}(a-d) shows that during the thermalization branch, the magnitude of heat released to the engineered cold reservoir increases with the thermalization duration while the erasure cost first increases rapidly and then tends to be stable. Meanwhile, the work is extracted through the compression-expansion stroke during the unitary branch. Fig.~\ref{fig2}(e,f), it further show that the engine starting from a nonstationary initial state converges to a stable cycle after several iterations. The branch probabilities evolve as $p^{(i+1)}=Pp^{(i)}$, with $P$ determined by the measurement basis and two feedback operations, and exponentially approach a stationary distribution, leading to fixed thermodynamic quantities and efficiency~\cite{SM}. The efficiency of the engine stabilizes at $59.1(1.6)\%$, consistent with the theoretical value of $59.9\%$ and the surpassing corresponding Otto benchmark $\eta_{\text{Otto}}=1-\omega_c/\omega_h$ by $18.2\%$. Meanwhile, this stable operation and resolved energetic balance provide the experimental foundation for further identifying the separate roles of measurement-induced coherence and inner friction.
	
	We then vary the measurement angle $\theta$ to demonstrate the efficiency advantage of stabilized MFQIE. In \autoref{fig3}(a,b), we control the angle $\theta$ of measurement basis to collapse the state of the system into different coherent eigenstates and observe two distinct regimes. For a small $\theta<0.04\pi$, the heat engine works as a conventional heat engine, outputting work by consuming energy provided by the heat absorption from the cold bath with the efficiency lower than corresponding Otto benchmark. However, as $\theta$ increases, the engine enters a measurement-enhanced regime with the efficiency beyond the corresponding Otto benchmark, where the measurement-induced energy input supports work extraction and the heat is released to the cold reservoir. These regimes originate from a competition among measurement-induced energy, inner friction and heat dissipation, where a sufficiently large measurement-induced energy input changes the energetic balance to achieve an efficiency above the corresponding Otto benchmark. 
	To quantify this effect, we further evaluate the measurement-induced coherence by the relative entropy of coherence as $C(\theta)=p_+C(\rho_{t_0^+})+p_-C(\rho_{t_0^{-}})$ with $C(\rho)=S(\rho_{\mathrm{diag}})-S(\rho)$~\cite{PhysRevLett.113.140401}, where the efficiency does not strictly follow the monotonic increase of $C(\theta)$, as shown in \autoref{fig3}(b), since varying the measurement angle not only changes the measurement-induced energy input and post-measurement coherence, but also modifies the subsequent state evolution, thus influencing inner friction and heat dissipation. A self-consistent theoretical comparison with a fully dephased reference cycle, in which the postmeasurement populations are preserved while the energy-basis coherence is removed and both engines are iterated to their own steady cycles, shows that measurement-induced coherence yields a higher steady-state work output and efficiency in the relevant parameter regime~\cite{SM}. \autoref{fig3}(c) shows that the efficiency increases with the thermalization duration and then saturates. This behavior results from the redistribution of the steady-cycle branch probabilities and the corresponding changes in the balance among extracted work, measurement-induced energy input, heat exchange, and erasure cost. 
	
	\begin{figure}[htbp]
	\centering
	\includegraphics[width=0.9\linewidth]{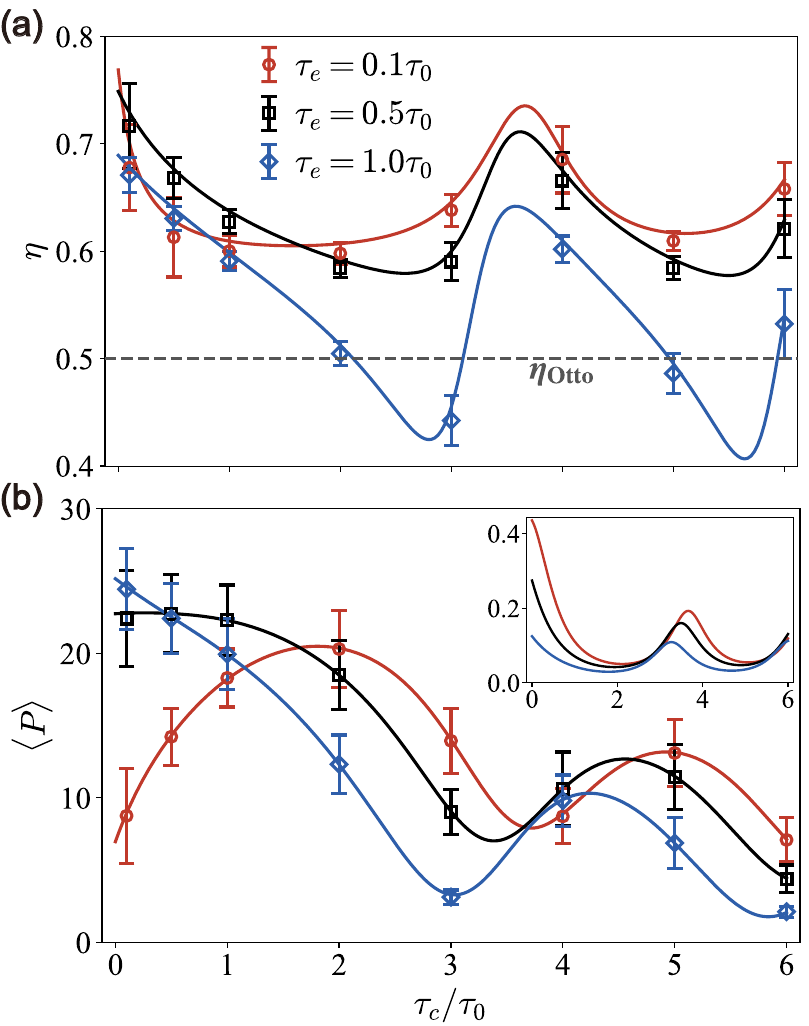}
	\caption{\textbf{Performance enhancement by quantum inner friction.} The efficiency \textbf{(a)} and power output \textbf{(b)} of MFQIE under different compression ($\tau_c$) and expansion durations ($\tau_e$) in the fully thermalized stable cycle, where the inset in (b) shows the inner-friction contribution to the extracted work. Here the intrinsic cycle power is defined as $\langle P\rangle = \langle w\rangle/\tau_{\mathrm{cycle}}$ with the total duration $\tau_{\mathrm{cycle}}=p_+(\tau_{c}+\tau_{e})+p_-\tau_{d}$, and the parameters are the same as those in \autoref{fig2}.}
	\label{fig4}
\end{figure}

	To identify the role of quantum inner friction, as shown in \autoref{fig4}, we experimentally measure the efficiency and power of the MFQIE by controlling the durations of finite-time unitary strokes. The MFQIE exhibits a duration-dependent damped-oscillatory behavior, where the nonnegative stroke-resolved inner-friction contribution to the extracted work varies with the stroke duration (see the inset of \autoref{fig4}(b)), enhancing the efficiency beyond the corresponding Otto benchmark and promoting the output power. In particular, at small $\tau_{c}$ the efficiency rapidly increases because of the larger positive inner-friction contribution, while the shorter cycle duration further increases the power; similarly, a shorter expansion stroke also tends to produce a higher efficiency. Meanwhile, the joint behavior of efficiency and power for different $\tau_e$ further shows that the performance is governed by the combined finite-time dynamics of the two unitary strokes rather than by either stroke alone. In \autoref{fig4}(b), the power is suppressed at very small $\tau_c$ for $\tau_e=0.1\tau_0$, since the reduction in extracted work outweighs the decrease in cycle duration, while the thermalization duration remains fixed. These results show that, in the present feedback cycle, nonadiabatic transitions need not act solely as a finite-time penalty. Over finite ranges of stroke durations, they allow the efficiency and intrinsic cycle power to increase simultaneously, thereby mitigating the usual finite-time efficiency-power compromise.

	\textit{Discussion and Conclusion\text{---}}The companion theoretical work develops the symmetric-FCS work quasiprobability, coherence-transition interference, correlated finite-cycle and fixed-time work statistics, and the entropy-rate reset cost of the outcome record for the same feedback architecture~\cite{Hong2026FeedbackEngine}. Another important extension is to implement real-time or autonomous feedback, where the measurement record is processed and reset within the same physical device, bringing the system closer to a complete Maxwell-demon cycle~\cite{PhysRevX.4.031015, PhysRevLett.110.040601, PhysRevX.7.021003,PRXQuantum.3.020329}. Further combining measurement-feedback control with a quantum load or quantum battery would connect information-to-work conversion with quantum energy storage~\cite{g45c-ssfx} and, together with resource-resolved analyses of coherence and finite-time nonadiabaticity~\cite{RevModPhys.89.041003, Lostaglio2015, Rezek_2006,PhysRevLett.113.260601, Liu_2025}, provide a route toward optimized quantum information engines with controllable efficiency, power, and irreversibility.
	
	In summary, we have experimentally demonstrated a measurement-feedback quantum information engine in which measurement-induced coherence and finite-time nonadiabatic dynamics can enhance work extraction. Using quantum-state tomography, we characterized the working-medium energy balance of the cycle, including extracted work, measurement-induced energy input, and heat exchange with the cold reservoir, while the specified conditional Landauer reset term is included in the operational efficiency accounting. The repeated feedback map converges to a stable operating regime whose efficiency can exceed the corresponding Otto benchmark. Over finite ranges of stroke durations, the finite-time nonadiabatic contribution allows the efficiency and intrinsic cycle power to increase simultaneously.

	\textit{Acknowledgments\text{---}}This work is supported by the National Key Research and Development Program of China under Grant No. 2022YFA1404500, by Cross-disciplinary Innovative Research Group Project of Henan Province under Grant No. 232300421004, National Natural Science Foundation of China under Grant Nos. 1232410, U24A2015, U21A20434, 12074346, 12274376, 12374466, 12074232, 12125406, 12504591, by Natural Science Foundation of Henan Province under Grant Nos. 232300421075, 242300421212, 252300421207, by Major science and technology project of Henan Province under Grant No. 221100210400, by the China Postdoctoral Science Foundation under Grants No. BX20250179. 
	\bibliography{REV}
	
\end{document}